\documentclass[]{aa}

\usepackage{longtable}
\usepackage{amsmath}
 
\usepackage{float}

\usepackage{tikz}
\usetikzlibrary{arrows.meta,positioning,decorations.markings}
\usetikzlibrary{arrows.meta,positioning}
\usetikzlibrary{arrows.meta,decorations.pathmorphing,positioning}
\usetikzlibrary{arrows.meta,positioning,decorations.pathreplacing}

\usepackage{graphicx}
\usepackage{subfigure}
\usepackage{adjustbox}
\usepackage{txfonts}
\usepackage{natbib}
\usepackage{hyperref}

\begin{document}

   \title{The different methods to calculate cluster membership probabilities}

  \author{T.~Ramezani\inst{1}
	 \and P.~Mondal\inst{1}
	 \and K.~Neumannov{\'a}\inst{1}
  \and E.~Paunzen\inst{1}
  \and J.~Sup{\'i}kov{\'a}\inst{1,2}
	 \and G.~Sz{\'a}sz\inst{1}}

   \institute{Department of Theoretical Physics and Astrophysics, Faculty of Science, Masaryk University, Kotl\'{a}\v{r}sk\'{a} 2, 611 37 Brno, Czechia
   \email{Taherehramezani7@gmail.com}
    \and CESNET, Gener\'{a}la Píky 430/26 160 00 Prague 6, Czechia}
   \date{}

\date{}
 
  \abstract
   {Reliable membership determination is a fundamental step in the study of star clusters. With the advent of $Gaia$ astrometry, a wide range of statistical and machine-learning techniques has been developed to assign membership probabilities. However, the current situation of membership lists is very unsatisfactory.}
   {This review summarises the main methodologies, compares their strengths and limitations, and discusses future directions. The aim is to provide a comprehensive overview and to lead to a more efficient and
   reliable approach for the forthcoming $Gaia$ DR4.}
   {Basically, we know of spatial, classical kinematic, and photometric methods, as well as maximum likelihood and Bayesian statistical methods, and machine learning and clustering algorithms. These different methods
   come with many modifications and flavours.}
   {We assessed all the advantages and disadvantages of the known methods to determine cluster 
   membership probabilities. Although nowadays most methods are based on poor statistical numerics,
   the more robust algorithms should still be taken into account. It is important to apply and compare
   several methods.}
   {The next step must be to define a list of standard star clusters to test and verify all known methods.
   The list must cover the complete grid of cluster parameters (age, distance, reddening, and metallicity)
   and total masses.}


   \maketitle
%

\section{Introduction} \label{introduction} 

Open clusters are among the most fundamental laboratories for studying stellar evolution, Galactic structure, and star formation. Because the member stars of an open cluster are believed to have formed nearly simultaneously from the same molecular cloud, they share similar ages, chemical compositions, and distances while exhibiting only small intrinsic variations in their physical properties. Consequently, open clusters provide indispensable benchmarks for calibrating stellar evolutionary models, determining Galactic abundance gradients, tracing spiral arm structure, and constraining the dynamical evolution of the Milky Way \citep{lynga1987,janes1988,portegies2010,cantatgaudin2020}.

One of the most important challenges in open-cluster research is the reliable identification of member stars. Every photometric or astrometric observation of a cluster inevitably includes contaminating foreground and background stars that are unrelated to the cluster. These field stars can significantly bias estimates of fundamental cluster parameters such as age, reddening, metallicity, distance modulus, luminosity function, present-day mass function, binary fraction, internal velocity dispersion, and tidal radius if they are not properly removed. Consequently, accurate determination of cluster membership probabilities has become one of the central problems in Galactic stellar astronomy.

Historically, cluster membership determination relied primarily on projected stellar positions and visual inspection of photographic plates. Early investigations assumed that stars concentrated toward the apparent cluster centre were likely members, while more distant stars were considered field contaminants \citep{trumpler1930}. Although such methods were useful for nearby rich clusters, they lacked statistical rigor and were often unreliable for sparse or heavily contaminated systems.

The first quantitative approaches emerged during the second half of the twentieth century with the introduction of proper-motion measurements. Since cluster members share a common space motion, their proper motions form a compact distribution that is distinguishable from the broader distribution of field stars. This concept led to the development of classical parametric methods, in which both cluster and field populations are represented by analytical probability density functions, typically multivariate Gaussian distributions. The pioneering work of \citet{sanders1971} established the maximum-likelihood framework that remains the basis of many modern membership algorithms. Subsequent improvements incorporated realistic measurement uncertainties, heteroscedastic errors, and correlated astrometric quantities \citep{stetson1987,zhao1990}.

The advent of large CCD surveys dramatically increased the availability of precise photometric measurements, enabling membership analyses based on colour--magnitude diagrams (CMDs). Since cluster members occupy well-defined evolutionary sequences, stars lying close to theoretical isochrones or empirical fiducial sequences are more likely to belong to the cluster than stars scattered throughout the CMD. Numerous studies, therefore, combined astrometric and photometric information to improve membership reliability, especially for distant clusters where proper-motion uncertainties remained significant \citep{kharchenko2004,kharchenko2013}.

A transformational breakthrough occurred with the launch of the $Gaia$ \citep{gaia2016} mission. The successive $Gaia$ Data Releases, particularly $Gaia$ DR2 \citep{gaia2018dr2} and $Gaia$ DR3 \citep{gaia2022dr3}, have provided unprecedented astrometric precision for more than one billion stars, including positions, parallaxes, proper motions, radial velocities, and homogeneous photometry. For nearby open clusters, $Gaia$ has reduced proper-motion uncertainties to a few tens of microarcseconds per year, making membership determination considerably more robust than ever before. These datasets have enabled comprehensive catalogues of Galactic open clusters containing homogeneous membership probabilities for thousands of systems \citep{cantatgaudin2018,cantatgaudin2020,castroginard2020,dias2021}.

In parallel with these observational advances, statistical methodologies have evolved substantially. Traditional maximum-likelihood estimators have been complemented by non-parametric kernel density estimation, Bayesian inference, Gaussian mixture models, expectation--maximization algorithms, hierarchical clustering, density-based clustering techniques such as DBSCAN and HDBSCAN, graph-based methods, and increasingly sophisticated machine-learning algorithms including Random Forests, Support Vector Machines, Artificial Neural Networks, and Deep Learning approaches \citep{krone2014,cantatgaudin2018}. Modern analyses frequently combine astrometric, photometric, spectroscopic, and spatial information simultaneously within multidimensional parameter spaces.

However, looking at the two most recent catalogues of star cluster members 
\citep{2023A&A...673A.114H,2023MNRAS.526.4107P}, we find very different lists for the same star cluster
although based on the $Gaia$ DR3. 

Each methodology possesses distinct strengths and limitations. Parametric methods generally provide statistically interpretable probabilities but rely on assumptions regarding the underlying distributions. Non-parametric techniques offer greater flexibility but often require careful bandwidth selection and can become computationally demanding for large datasets. Bayesian approaches naturally incorporate prior information and observational uncertainties but may depend sensitively on the adopted priors. Machine-learning methods excel at identifying complex non-linear decision boundaries but usually require representative training samples and may sacrifice interpretability for predictive performance.

The rapidly increasing volume of astronomical survey data presents additional challenges. Future facilities such as the Vera C. Rubin Observatory Legacy Survey of Space and Time (LSST), the Nancy Grace Roman Space Telescope, and future $Gaia$ releases will produce petabyte-scale datasets that require highly automated, scalable, and statistically robust membership-determination techniques. Consequently, developing algorithms capable of handling multidimensional, heterogeneous, and incomplete data has become one of the foremost objectives in Galactic astronomy.

The purpose of this review is to provide a comprehensive overview of the methodologies developed for determining stellar membership probabilities in Galactic open clusters. We summarise the historical evolution of membership analysis, review the principal statistical and machine-learning approaches currently employed in the literature, compare their assumptions, computational characteristics, advantages, and limitations, and discuss future directions motivated by next-generation astronomical surveys. Rather than advocating a single optimal technique, this review aims to provide researchers with a unified framework for understanding the diverse methodologies available and for selecting the most appropriate approach according to the characteristics of their observational data.

\section{Overview of the literature} \label{overview}

First, we want to give a comprehensive overview of the literature which dealt with the determination
of membership probabilities of star clusters. 

\subsection{The Birth of Cluster Membership Studies (1930--1957)}

The first systematic attempts to identify members of open clusters emerged in the early twentieth century, driven by the need to understand the spatial distribution and physical reality of stellar groupings in the Milky Way. Early work was primarily observational and descriptive, relying on star counts, visual inspection of photographic plates, and rudimentary astrometric measurements. Despite their simplicity, these studies laid the foundation for later statistically rigorous approaches.

A key milestone was the work of \citet{trumpler1930}, who provided the first quantitative evidence that open clusters are physically coherent systems rather than chance alignments of stars. By analysing photographic observations of several clusters, Trumpler demonstrated systematic trends in cluster diameters, distances, and spatial distribution. Importantly, he also identified the effects of interstellar extinction, showing that clusters appeared dimmer and more distant than previously assumed. This result had profound implications: it established that open clusters could be used as probes of Galactic structure, but only if their true stellar populations could be reliably identified.

During this early period, membership determination was not yet formalised probabilistically. Instead, researchers relied on spatial concentration criteria, assuming that stars near the apparent cluster centre were more likely to be members. This assumption can be expressed in its simplest form as:
\begin{equation}
P_{\mathrm{mem}} \propto \Sigma(r),
\end{equation}
where $\Sigma(r)$ is the projected surface density of stars as a function of radial distance from the cluster centre. However, this approach is strongly biased by field-star contamination, particularly in low-density clusters or regions near the Galactic plane.

Early attempts to improve membership identification included star counts in concentric annuli and comparisons with adjacent control fields. While these methods provided statistical corrections for background contamination, they did not account for kinematic or photometric information and, therefore, could not distinguish overlapping stellar populations with similar spatial distributions.

By the mid-twentieth century, advances in photographic astrometry enabled more precise measurements of stellar positions over long time baselines. This development enabled astronomers to begin exploring stellar motions as a criterion for membership. The fundamental idea was that cluster members share a common space motion, leading to coherent proper-motion patterns on the sky. Although this concept would later become central to modern membership analysis, its early application remained qualitative.

A representative example of this transition is the early work on nearby clusters such as the Pleiades and Hyades, where multi-epoch photographic plates were used to identify stars with common apparent motion. These studies demonstrated that clusters are dynamically coherent systems, but measurement uncertainties were still too large to derive robust membership probabilities.

The observational limitations of this era can be summarised as follows:
(i) photographic plate distortions and systematic errors,
(ii) limited time baselines (often insufficient for precise proper motions),
(iii) lack of homogeneous photometric systems,
(iv) absence of statistical frameworks for population separation.

Despite these limitations, this period introduced the conceptual basis for all later methods: the recognition that cluster membership must be treated as a statistical separation problem between overlapping stellar populations.

A summary of the main characteristics of early membership approaches is provided in Table~\ref{tab:early_methods}.

The conceptual evolution of this period can be visualised as a progression from purely spatial reasoning toward kinematic interpretation, even though a fully statistical framework had not yet been established.

By the end of this era (circa 1957), it had become evident that spatial information alone was insufficient for robust membership determination. This realisation directly motivated the development of formal kinematic models in the following decade, culminating in the introduction of statistical proper-motion membership methods.

\subsubsection{First Photographic Investigations}

The introduction of long-exposure photographic plates marked a turning point in early cluster studies, enabling the first systematic surveys of stellar positions across extended fields. Unlike visual star counts, photographic techniques enabled astronomers to record faint stars and measure their relative positions with greater precision, though they were still limited by emulsion distortions and plate-to-plate inconsistencies.

Early photographic investigations primarily focused on nearby and well-known clusters such as the Pleiades, Hyades, and Praesepe. These systems were particularly suitable because their angular sizes were large and their member stars relatively bright. Multi-epoch observations enabled astronomers to compare stellar positions over time intervals of several years to decades, providing the first qualitative estimates of stellar motion.

A key limitation of these early studies was the lack of a rigorous statistical framework. Membership assignment was typically based on a combination of:
(i) proximity to the cluster centre,
(ii) similarity in brightness and colour,
(iii) apparent consistency in motion across photographic epochs.

Despite these limitations, photographic studies provided the first empirical evidence that open clusters contain dynamically coherent stellar populations embedded within dense Galactic fields. However, contamination by field stars remained a fundamental challenge, particularly in low-latitude clusters.

\subsubsection{The Proper-Motion Transition (1958--1965)}

The late 1950s and early 1960s represent a fundamental turning point in cluster membership studies, as the field transitioned from purely spatial and photographic approaches toward quantitative kinematic separation. This period established the first statistical frameworks capable of distinguishing cluster members from field stars using proper-motion information.

A key development was the recognition that stars belonging to an open cluster share a common space motion, which projects into a compact distribution in the proper-motion plane. In contrast, field stars exhibit a broader and more dispersed kinematic distribution due to the superposition of multiple Galactic components. This insight allowed membership determination to be formulated as a statistical classification problem.

\citet{vasilevskis1958} were among the first to formalise this idea using photographic proper-motion measurements. Their work introduced the concept that the observed proper-motion distribution can be decomposed into two overlapping populations: a concentrated cluster component and a broad field component. Although the original formulation was not yet fully probabilistic in the modern Bayesian sense, it represented a decisive step toward statistical membership analysis.

In its simplest form, the proper-motion distribution can be written as a mixture model:
\begin{equation}
\Phi(\mu) = n_c \, \Phi_c(\mu) + n_f \, \Phi_f(\mu),
\end{equation}
where $\Phi_c(\mu)$ and $\Phi_f(\mu)$ represent the cluster and field proper-motion distributions, respectively, and $n_c$, $n_f$ are their normalised fractions.

A common assumption in early studies was that the cluster distribution can be approximated by a bivariate Gaussian:
\begin{equation}
\Phi_c(\mu) = \frac{1}{2\pi \sigma^2} \exp\left[-\frac{(\mu_x - \mu_{x,c})^2 + (\mu_y - \mu_{y,c})^2}{2\sigma^2}\right],
\end{equation}
where $(\mu_{x,c}, \mu_{y,c})$ represents the mean cluster motion and $\sigma$ the intrinsic dispersion combined with observational uncertainties.

The field population was often modelled empirically as a broader, asymmetric distribution, sometimes approximated by a Gaussian with different dispersions along the two axes or even by exponential functions in one dimension. Despite these simplifications, the separation of the two populations in proper-motion space became a powerful diagnostic tool.

The conceptual advance of this period lies in the implicit introduction of probability as a criterion of membership. Instead of assigning membership based on deterministic thresholds, stars could now be evaluated according to their likelihood of belonging to the cluster:
\begin{equation}
P_{\mathrm{mem}} = \frac{n_c \, \Phi_c(\mu)}{n_c \, \Phi_c(\mu) + n_f \, \Phi_f(\mu)}.
\end{equation}

Although this expression was not yet widely standardised in the 1960s, it forms the basis of later formalisations, including the maximum-likelihood framework developed in the following decade.

\citet{brosche1989} summarised the refined methods, which incorporated improved plate reductions and longer time baselines, thereby significantly reducing proper-motion uncertainties. These improvements allowed clearer separation between cluster and field populations, particularly for nearby clusters such as the Hyades.

The advantages of proper-motion-based membership determination over purely spatial methods can be summarised in Table~\ref{tab:pm_advantages}.

By the mid-1960s, these developments had clearly demonstrated that kinematic information provides a far more robust discriminator of cluster membership than spatial data alone. However, the methods still relied on simplifying assumptions about Gaussianity and did not fully incorporate measurement uncertainties within a rigorous statistical framework. These limitations directly motivated the development of maximum-likelihood methods in the following decade, most notably the seminal work of \citet{sanders1971}, which formalised membership probabilities in a consistent statistical framework.

\subsubsection{Toward Modern Statistical Methods (1965--1970)}

By the late 1960s, cluster membership studies had reached a critical conceptual transition. While earlier approaches established that proper-motion space can separate cluster and field populations, the methods remained largely heuristic and dependent on simplified assumptions. The period between 1965 and 1970 marks the emergence of a more rigorous statistical perspective, in which membership determination began to be formulated explicitly as a likelihood maximisation problem.

A central limitation of earlier work was the lack of a unified probabilistic framework that consistently incorporated observational uncertainties, intrinsic dispersions, and overlapping distributions. This motivated the development of statistical approaches in which each star is assigned a probability of membership based on the relative likelihood of its observed properties under competing population models.

The core idea can be expressed in a general form as:
\begin{equation}
P_{\mathrm{mem}, i} = \frac{\mathcal{L}_c(\mathbf{x}_i)\, \pi_c}{\mathcal{L}_c(\mathbf{x}_i)\, \pi_c + \mathcal{L}_f(\mathbf{x}_i)\, \pi_f},
\end{equation}
where $\mathbf{x}_i$ represents the observed properties of star $i$ (typically proper motion components, and in some cases magnitudes), $\mathcal{L}_c$ and $\mathcal{L}_f$ are the likelihood functions for cluster and field populations, and $\pi_c$, $\pi_f$ are prior probabilities.

This formulation represents a conceptual precursor to modern Bayesian inference, even though it was not yet fully formalised in that language. The key advancement was the explicit recognition that membership is inherently probabilistic and that uncertainties must be incorporated into the model rather than treated as post-processing corrections.

During this period, several authors refined the modelling of proper-motion distributions. The cluster population was increasingly treated as a Gaussian distribution with intrinsic dispersion, while the field population required more flexible representations due to its composite Galactic nature. Some studies introduced anisotropic Gaussian models for the field, allowing different dispersions along Galactic longitude and latitude directions.

The observational improvements in this era were also significant. Longer temporal baselines between photographic plates led to improved proper-motion precision, reducing statistical overlap between cluster and field populations. This improvement was particularly important for nearby clusters, where internal velocity dispersions could begin to be marginally resolved.

Despite these advances, a number of limitations remained:

\begin{itemize}
\item The assumption of Gaussianity for both cluster and field populations was often unrealistic.
\item Measurement errors were not always explicitly propagated into likelihood functions.
\item Photometric information was rarely incorporated in a self-consistent statistical framework.
\item Membership probabilities were still sensitive to subjective choices in model parameterisation.
\end{itemize}

Nevertheless, the methodological advances of this period directly paved the way for the seminal work of \citet{sanders1971}, who provided the first widely adopted maximum-likelihood framework for cluster membership determination. Sanders' formulation unified earlier ideas into a consistent statistical model and remains a foundational reference for modern approaches.

A comparison of methodological characteristics across this transitional period is summarised in Table~\ref{tab:transition_ml}.

By the end of this period, the conceptual foundation of modern cluster membership analysis had been established. The recognition that stellar membership must be inferred statistically, rather than assigned deterministically, represents a fundamental shift in methodology. This paradigm directly enabled the development of maximum-likelihood methods in the 1970s and, much later, the Bayesian and machine-learning approaches used in contemporary surveys such as Gaia.

\subsection{The Maximum-Likelihood Era (1971--1990)}

The introduction of a rigorous maximum-likelihood formalism by \citet{sanders1971} marked the beginning of modern statistical approaches to cluster membership determination. In contrast to earlier heuristic or semi-empirical methods, this framework explicitly treated cluster membership as a statistical inference problem, in which the observed proper-motion distribution is modelled as a mixture of cluster and field populations.

In Sanders' formulation, the likelihood of observing a star with proper motion $\boldsymbol{\mu}_i = (\mu_{x,i}, \mu_{y,i})$ is given by:
\begin{equation}
\mathcal{L}_i = n_c \, \Phi_c(\boldsymbol{\mu}_i) + n_f \, \Phi_f(\boldsymbol{\mu}_i),
\end{equation}
where $\Phi_c$ and $\Phi_f$ denote the probability density functions of the cluster and field populations, respectively, and $n_c$ and $n_f$ are their normalised contributions.

The cluster component is typically modelled as a bivariate Gaussian distribution:
\begin{equation}
\Phi_c(\boldsymbol{\mu}) =
\frac{1}{2\pi \sigma_c^2}
\exp\left[
-\frac{(\mu_x - \mu_{x,c})^2 + (\mu_y - \mu_{y,c})^2}{2\sigma_c^2}
\right].
\end{equation}

The corresponding membership probability for each star is then expressed as:
\begin{equation}
P_{\mathrm{mem}, i} =
\frac{n_c \, \Phi_c(\boldsymbol{\mu}_i)}
{n_c \, \Phi_c(\boldsymbol{\mu}_i) + n_f \, \Phi_f(\boldsymbol{\mu}_i)}.
\end{equation}

This formulation represents the first widely adopted probabilistic framework for cluster membership and remains conceptually central to many modern algorithms.

Following Sanders' work, the 1970s and 1980s saw a series of refinements aimed at improving both the realism of the adopted models and the robustness of the parameter estimation. A major development was the inclusion of iterative procedures for estimating cluster parameters (mean motion, dispersion, and field contamination fraction), often solved via expectation-like schemes even before formal EM algorithms were widely adopted in astronomy.

\citet{stetson1987} and subsequent studies emphasized the importance of accounting for observational uncertainties and crowding effects in dense stellar fields. Although their work was primarily focused on photometric and astrometric reduction techniques, it contributed indirectly to membership studies by improving the quality of input data.

During the same period, \citet{zhao1990} and related works extended the maximum-likelihood framework by introducing more flexible representations of the field-star distribution, often using asymmetric or multi-component Gaussian models to better reflect the complexity of Galactic kinematics near the solar neighbourhood.

A key limitation of this era remained the assumption of relatively simple functional forms for both cluster and field distributions. While Gaussian models are mathematically convenient, real stellar populations often exhibit non-Gaussian features due to differential Galactic rotation, moving groups, and observational selection effects.

Despite these limitations, the maximum-likelihood framework provided three fundamental advances:

\begin{itemize}
\item A formal probabilistic definition of membership.
\item A self-consistent way to estimate cluster parameters and membership simultaneously.
\item A scalable framework applicable to large stellar samples.
\end{itemize}

The methodological landscape of this period can be summarised in Table~\ref{tab:ml_1970_1990}.

This period established the conceptual and mathematical foundation for all subsequent membership-determination methods. In particular, the idea that cluster membership can be expressed as a normalised likelihood ratio remains central to modern Bayesian approaches and machine-learning classifiers.

The transition from this classical era toward multi-dimensional and survey-driven approaches began in the late 1980s, as CCD photometry and larger astrometric catalogues started to replace photographic plate data. This transition is addressed in the following subsection.

\subsection{The Photometric--Astrometric Expansion (1990--2010)}

The period from the early 1990s to the end of the 2000s marks a major expansion in cluster membership studies, driven by the widespread availability of CCD photometry and improved astrometric catalogues such as Hipparcos and early ground-based surveys. This era is characterised by the transition from purely kinematic models to multi-dimensional approaches combining astrometry, photometry, and spatial information.

A key limitation of classical maximum-likelihood methods was their reliance solely on proper-motion space, which becomes insufficient in regions where cluster and field kinematics overlap. To overcome this degeneracy, researchers began incorporating photometric constraints, particularly colour--magnitude diagram (CMD) information, into membership estimation.

One of the earliest and most influential ideas was that cluster members should lie along a well-defined isochrone in the CMD, while field stars are broadly distributed. This led to the development of hybrid methods combining spatial density profiles with photometric filtering. In these approaches, the likelihood of membership can be generalised as:
\begin{equation}
\mathcal{L}_i = \mathcal{L}_{\mu}(\boldsymbol{\mu}_i)\,\mathcal{L}_{\mathrm{CMD}}(G_i, BP-RP_i)\,\mathcal{L}_{r}(r_i),
\end{equation}
where each term encodes information from proper motion, photometry, and spatial position, respectively.

During this period, several methodological families emerged.

\subsubsection{CMD-based membership filtering}

CMD-based approaches assume that cluster members occupy a narrow evolutionary sequence. Early implementations used manual or semi-automated ridge-line selection, which later evolved into statistical filtering techniques that quantify a star's distance from an empirical or theoretical isochrone. The probability can be expressed as:
\begin{equation}
P_{\mathrm{CMD}} \propto \exp\left(-\frac{d_{\mathrm{iso}}^2}{2\sigma_{\mathrm{phot}}^2}\right),
\end{equation}
where $d_{\mathrm{iso}}$ is the distance to the cluster isochrone in color--magnitude space.

\subsubsection{Spatial density and radial models}

In parallel, spatial structure models were refined using King-like density profiles and radial density functions. The projected stellar density was often modelled as:
\begin{equation}
\Sigma(r) = \frac{\Sigma_0}{1 + (r/r_c)^2},
\end{equation}
where $r_c$ is the core radius. This spatial information was combined with kinematic and photometric constraints to reduce field contamination.

\subsubsection{Kernel-based and non-parametric approaches}

A major methodological innovation of this era was the introduction of non-parametric density estimation techniques. Kernel density estimation (KDE) allowed the distribution of field stars in proper-motion or CMD space to be modelled without assuming Gaussianity. This was particularly important for crowded Galactic fields where multiple overlapping populations exist.

The general KDE form is:
\begin{equation}
\hat{f}(x) = \frac{1}{Nh} \sum_{i=1}^{N} K\left(\frac{x-x_i}{h}\right),
\end{equation}
where $h$ is the bandwidth parameter and $K$ is the kernel function.

\subsubsection{Early multi-dimensional membership frameworks}

By the mid-2000s, membership studies increasingly adopted multi-dimensional parameter spaces combining:
\[
(\mu_\alpha, \mu_\delta, V, (B-V), r),
\]
and later Gaia-like photometric systems. These approaches significantly improved membership reliability by reducing degeneracies between cluster and field populations.

Notable contributions include probabilistic decontamination techniques and statistical subtraction methods applied to CMDs of open clusters in dense Galactic regions. These methods typically involve constructing a reference field, CMD, and subtracting its statistical contribution from the cluster region.

\citet{bonatto2007} introduced a widely used statistical field-star decontamination algorithm that significantly improved the identification of cluster sequences in photometric diagrams. Similarly, \citet{balaguer2004} developed probabilistic membership assignments using combined astrometric and photometric criteria.

\subsubsection{Summary of methodological evolution}

The key advancement of this era was the transition from one-dimensional kinematic classification to multi-dimensional statistical inference. However, most methods still relied on heuristic combinations of independent probabilities rather than a fully unified probabilistic model.

A summary of the main methodological developments is provided in Table~\ref{tab:ccd_era_methods}.

\subsection{The $Gaia$ Revolution and High-Dimensional Membership (2016--present)}

The launch of the \textit{Gaia} mission marked a fundamental transformation in cluster membership studies. With the advent of $Gaia$ Data Release 2 \citep{gaia2018dr2} and Data Release 3 \citep{gaia2022dr3}, the astrometric precision improved by several orders of magnitude compared to pre-Gaia catalogues, enabling systematic membership determination for thousands of open clusters across the Milky Way.

For the first time, cluster membership could be studied in a nearly complete phase space, combining positions, proper motions, and parallaxes, and in some cases radial velocities. The effective dimensionality of the problem increased from 2D (proper motion) or 3D (proper motion + photometry) to a 5D or 6D space:
\begin{equation}
\mathbf{X} = (\alpha, \delta, \varpi, \mu_\alpha, \mu_\delta, v_r),
\end{equation}
where available.

This high-dimensional dataset enabled the development of fully data-driven clustering and probabilistic algorithms, significantly reducing the dependence on simplistic Gaussian assumptions.

\subsubsection{Gaussian Mixture and Bayesian approaches}

One of the most widely used modern frameworks is the Gaussian Mixture Model (GMM), in which the observed distribution is represented as a sum of multiple Gaussian components:
\begin{equation}
p(\mathbf{X}) = \sum_{k=1}^{K} \pi_k \, \mathcal{N}(\mathbf{X}|\boldsymbol{\mu}_k, \Sigma_k),
\end{equation}
where $\pi_k$ are mixture weights. Membership probabilities are then derived via posterior inference.

Bayesian approaches further extend this framework by explicitly incorporating priors on cluster structure, spatial distribution, and field-star contamination. These methods naturally propagate observational uncertainties and allow hierarchical modelling of cluster populations.

\subsubsection{UPMASK and non-parametric clustering}

A major milestone in this era is the development of the UPMASK algorithm \citep{krone2014}, which introduced a fully non-parametric, unsupervised approach to cluster membership determination. Instead of assuming parametric distributions, UPMASK operates by:

\begin{itemize}
\item Clustering stars in photometric or astrometric space using k-means,
\item Evaluating the spatial concentration of each group,
\item Iteratively resampling to estimate membership probabilities.
\end{itemize}

This method is particularly powerful for young or sparse clusters where classical Gaussian assumptions fail.

\subsubsection{Density-based clustering: DBSCAN and HDBSCAN}

Density-based methods such as DBSCAN and HDBSCAN became increasingly popular for identifying clusters in $Gaia$ data. These methods define clusters as phase-space overdensities without requiring prior assumptions about their shape.

In DBSCAN, a star is classified based on local density:
\begin{itemize}
\item Core points: high-density regions,
\item Border points: edge of clusters,
\item Noise points: field stars.
\end{itemize}

HDBSCAN extends this approach by constructing a hierarchical clustering structure, enabling more robust identification of variable-density systems.

\subsubsection{Large-scale $Gaia$ cluster catalogues}

The $Gaia$ era has enabled the construction of homogeneous membership catalogues for thousands of clusters. Among the most influential works are those by Cantat-Gaudin and collaborators, who systematically derived membership probabilities using astrometric clustering combined with photometric filtering.

These catalogues demonstrated that many previously known clusters were either:
(i) contaminated by field stars,
(ii) physically unbound associations,
or (iii) parts of larger stellar complexes.

Similarly, large surveys such as those by \citet{dias2021} and \citet{castroginard2020} expanded the census of Galactic clusters by applying automated pipelines to $Gaia$ data.

\subsubsection{Machine learning approaches}

Recent years have seen increasing adoption of machine learning techniques for membership determination. These include:

\begin{itemize}
\item Random Forest classifiers trained on confirmed members,
\item Support Vector Machines (SVMs) for nonlinear separation,
\item Neural networks for high-dimensional classification,
\item XGBoost and gradient boosting methods for probabilistic prediction.
\end{itemize}

These methods are particularly effective when training sets are available but may suffer from biases introduced by incomplete or non-representative training samples.

\subsubsection{Current challenges in the $Gaia$ era}

Despite the unprecedented precision of $Gaia$ data, several challenges remain:

\begin{itemize}
\item Non-Gaussian and asymmetric field-star distributions,
\item Unresolved binary contamination,
\item Differential extinction effects,
\item Degeneracies in sparse clusters,
\item Selection biases in radial velocity subsamples.
\end{itemize}

Moreover, the transition from classical probabilistic methods to machine learning has raised questions regarding interpretability and physical consistency.

\subsubsection{Summary of modern methods}

A condensed comparison of Gaia-era membership techniques is presented in Table~\ref{tab:gaia_methods}.

\section{Comparison of the literature} \label{comparison}

\subsection{Taxonomy of Membership Determination Methods}

The methods developed for cluster membership determination over the past decades can be broadly classified into five major families:

\begin{enumerate}
\item \textbf{Spatial methods} (1930--1950): Based solely on projected stellar density.
\item \textbf{Kinematic methods} (1950--1990): Based on proper-motion and radial-velocity separation.
\item \textbf{Photometric methods} (1990--2010): Based on CMD proximity to isochrones.
\item \textbf{Statistical probabilistic methods} (1971--present): Including maximum likelihood and Bayesian inference.
\item \textbf{Machine-learning and clustering methods} (2010--present): including supervised and unsupervised algorithms.
\end{enumerate}

Despite their differences, all approaches aim to estimate the posterior probability:
\begin{equation}
P(\mathrm{member} \mid \mathbf{X}) = \frac{P(\mathbf{X} \mid \mathrm{member}) P(\mathrm{member})}{P(\mathbf{X})},
\end{equation}
where $\mathbf{X}$ represents the observed multi-dimensional stellar parameters.

The main distinction between methods lies in how the likelihood term $P(\mathbf{X} \mid \mathrm{member})$ is modelled and which assumptions are imposed.

\subsection{Comparative Evaluation of Method Families}

A qualitative comparison of the principal methodological families is presented in Table~\ref{tab:method_families}.

\subsection{Discussion of Methodological Assumptions}

A central aspect in the comparison of membership methods is the underlying assumption regarding the distribution of field stars. Early methods assumed either uniform spatial distributions or simple Gaussian kinematic profiles. However, $Gaia$ data have demonstrated that field populations are highly structured, containing moving groups, tidal streams, and asymmetric velocity distributions.

This complexity violates the assumptions of classical maximum-likelihood methods, which has motivated the adoption of nonparametric and machine-learning approaches in recent years.

Another key distinction lies in dimensionality. Earlier methods typically operated in two-dimensional proper-motion space, whereas modern approaches exploit up to six dimensions, including parallax and radial velocity. Increasing dimensionality generally improves classification performance but introduces challenges related to correlated uncertainties and missing data.

\subsection{Bias--Variance Trade-off in Membership Methods}

Membership determination methods can also be interpreted in terms of the bias--variance trade-off:

\begin{itemize}
\item \textbf{Low-complexity models} (e.g., Gaussian mixtures) exhibit high bias but low variance.
\item \textbf{Non-parametric methods} reduce bias but increase variance.
\item \textbf{Machine learning models} attempt to balance both through regularisation and training data.
\end{itemize}

This trade-off becomes particularly important in sparse clusters, where overfitting field structures can significantly bias membership probabilities.

\subsection{Dimensionality versus Reliability}

A general trend observed across the literature is that membership reliability improves with the inclusion of additional independent observables:
\begin{equation}
\mathrm{Reliability} \propto f(N_{\mathrm{dimensions}}),
\end{equation}
although this relation saturates when systematic uncertainties dominate over statistical ones.

Gaia has effectively shifted this regime from low-dimensional statistical separation to high-dimensional structured inference.

\subsection{Performance Trends Across Method Families}

Beyond methodological classification, a key question is how different approaches perform as a function of data dimensionality, cluster richness, and field contamination. While a fully quantitative comparison is difficult due to heterogeneous datasets in the literature, general trends can be identified.

A consistent result is that methods incorporating higher-dimensional information (astrometry + photometry + spectroscopy) achieve significantly improved membership reliability compared to single-domain approaches. However, this improvement saturates when systematic uncertainties dominate.

This trend can be schematically represented as an increasing but saturating function of dimensionality:
\begin{equation}
R(N_d) \sim R_{\infty} \left(1 - e^{-\alpha N_d}\right),
\end{equation}
where $R$ denotes classification reliability and $N_d$ the number of independent dimensions.

\subsection{Method Performance in Different Cluster Environments}

The performance of membership methods depends strongly on the cluster environment:

\begin{itemize}
\item \textbf{Rich nearby clusters:} Kinematic and Gaia-based methods perform best due to clear phase-space separation.
\item \textbf{Sparse or dissolving clusters:} Density-based and machine-learning methods outperform Gaussian models.
\item \textbf{Highly reddened regions:} Photometric methods become unreliable without accurate extinction correction.
\item \textbf{Dense Galactic fields:} Non-parametric and clustering algorithms are required to disentangle overlapping populations.
\end{itemize}

\subsection{Decision Framework for Method Selection}

Given the diversity of available approaches, a practical decision framework is essential. The choice of method depends primarily on data dimensionality, cluster density, and the availability of training labels.

\subsection{Global Comparison of Method Performance}

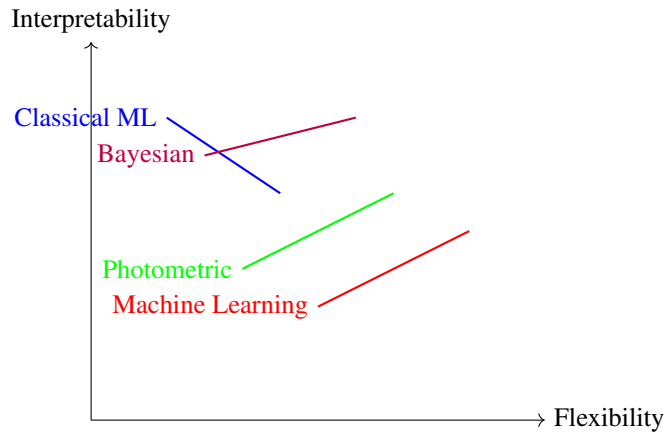
\begin{figure*}
\centering

\begin{tikzpicture}

\draw[->] (0,0) -- (6,0) node[right] {Flexibility};
\draw[->] (0,0) -- (0,5) node[above] {Interpretability};

\draw[blue, thick] (1,4) node[left] {Classical ML} -- (2.5,3);
\draw[green, thick] (2,2) node[left] {Photometric} -- (4,3);
\draw[red, thick] (3,1.5) node[left] {Machine Learning} -- (5,2.5);
\draw[purple, thick] (1.5,3.5) node[left] {Bayesian} -- (3.5,4);

\end{tikzpicture}

\caption{Qualitative comparison of membership determination methods in terms of interpretability versus flexibility. Classical methods are highly interpretable but less flexible, while machine-learning approaches show the opposite trend.}
\label{fig:performance_trend}
\end{figure*}

A schematic comparison of the relative performance of different method families is shown in Fig.~\ref{fig:performance_trend}. The figure illustrates the trade-off between interpretability and flexibility across methodological classes.

\subsection{Critical Limitations Across All Methods}

Despite major progress, several limitations persist across all methodological families:

\begin{itemize}
\item \textbf{Field-star complexity:} Real Galactic populations deviate significantly from simple analytic models.
\item \textbf{Binary contamination:} Unresolved binaries bias photometric and astrometric distributions.
\item \textbf{Extinction variability:} Differential reddening affects CMD-based methods.
\item \textbf{Selection effects:} Survey incompleteness introduces systematic biases.
\item \textbf{Model dependence:} Probabilistic outputs depend strongly on assumed priors or training data.
\end{itemize}

These limitations highlight that membership determination remains an intrinsically ill-posed inverse problem, even with high-quality $Gaia$ data.

\subsection{Synthesis}

Overall, the literature demonstrates a clear evolutionary trend:

\begin{center}
Spatial $\rightarrow$ Kinematic $\rightarrow$ Photometric $\rightarrow$ Probabilistic $\rightarrow$ Machine Learning
\end{center}

Each stage represents an increase in dimensionality, statistical rigour, and computational complexity. However, no single method is universally optimal; instead, method selection must be guided by data quality, cluster environment, and scientific objectives.

\section{Future Prospects}

The next decade of cluster membership studies will be shaped by both upcoming astronomical surveys and advances in statistical and machine-learning methodologies. While $Gaia$ has already transformed the field, several fundamental limitations remain, particularly in crowded regions, faint regimes, and highly dynamic environments such as dissolving clusters and stellar streams.

\subsection{Next-generation observational surveys}

Future surveys will significantly extend both the depth and dimensionality of available data.

The Vera C. Rubin Observatory Legacy Survey of Space and Time (LSST) will provide deep, time-resolved photometry over a large fraction of the sky. This will enable variability-based membership constraints and improved separation of field contaminants through long-term photometric behaviour.

Similarly, the Nancy Grace Roman Space Telescope will deliver high-precision near-infrared astrometry and photometry in crowded regions of the Galactic bulge and disk, where $Gaia$ performance is limited by extinction and crowding.

In addition, future $Gaia$ data releases (DR4 and DR5) are expected to improve radial velocity coverage, orbital solutions, and long-term astrometric precision, enabling more complete 6D phase-space reconstruction for large samples of clusters.

\subsection{From static membership to dynamical membership}

A major conceptual shift in future studies will be the transition from static membership classification to dynamical membership inference. Instead of treating clusters as instantaneous overdensities in phase space, future models will incorporate orbital evolution, tidal stripping, and time-dependent phase-space distributions.

In this framework, membership becomes a function of time:
\begin{equation}
P_{\mathrm{mem}} = P_{\mathrm{mem}}(\mathbf{X}, t),
\end{equation}
where stellar positions are modelled within dynamical potentials rather than static distributions.

This is particularly relevant for dissolving clusters and tidal debris, where classical clustering assumptions break down.

\subsection{Machine learning and artificial intelligence}

Modern machine learning techniques are expected to play an increasingly important role in membership determination. However, future progress will likely focus not only on predictive accuracy but also on interpretability and physical consistency.

Emerging approaches include:

\begin{itemize}
\item \textbf{Physics-informed neural networks (PINNs):} embedding dynamical constraints into learning architectures.
\item \textbf{Graph neural networks (GNNs):} modeling stellar systems as relational structures in phase space.
\item \textbf{Probabilistic deep learning:} providing calibrated uncertainties rather than deterministic labels.
\item \textbf{Diffusion models:} reconstructing underlying cluster distributions from noisy observations.
\end{itemize}

These methods aim to bridge the gap between purely data-driven classification and physically motivated modelling.

\subsection{Explainability and uncertainty quantification}

A critical requirement for future methods is interpretability. While deep learning models can achieve high classification accuracy, their lack of transparency poses challenges for physical interpretation.

Future frameworks will therefore emphasise:

\begin{itemize}
\item Uncertainty calibration,
\item Robust error propagation,
\item Model interpretability,
\item Reproducibility across surveys.
\end{itemize}

This is particularly important in astrophysical contexts, where membership probabilities directly impact derived cluster parameters such as age, metallicity, and mass functions.

\subsection{Towards unified probabilistic frameworks}

The long-term goal of cluster membership studies is the development of a fully unified probabilistic framework that integrates:

Astrometry, photometry, spectroscopy, time-domain, and dynamics.

Such a framework would treat membership inference as a hierarchical Bayesian problem, in which clusters are modelled as evolving dynamical systems embedded in a structured Galactic potential.

\subsection{Summary outlook}

In summary, future advances will be driven by three converging factors:

\begin{itemize}
\item Deeper and more precise multi-survey data,
\item Increasing dimensionality and time-domain coverage,
\item More sophisticated probabilistic and machine-learning models.
\end{itemize}

Together, these developments will move cluster membership studies from static classification problems toward fully dynamical, data-driven inference of stellar populations in the Milky Way.

\section{Conclusions} \label{conclusions}

This review provides a comprehensive overview of methods for determining membership probabilities in Galactic open clusters, tracing their evolution from early spatial techniques to modern machine-learning and Gaia-driven high-dimensional frameworks.

The historical development reveals a clear progression in both conceptual understanding and methodological sophistication. Early studies relied exclusively on spatial overdensities, assuming that clusters could be identified through simple star-count enhancements. While these approaches provided the first evidence for the physical reality of open clusters, they were fundamentally limited by severe field-star contamination and the absence of kinematic information.

The introduction of photographic astrometry and proper-motion measurements marked the first major transition toward physically motivated membership criteria. This led to the development of bivariate kinematic separation methods and ultimately to the formalisation of statistical membership inference through maximum-likelihood approaches. The seminal work of \citet{sanders1971} established a rigorous probabilistic framework that remains conceptually central to modern methods.

Subsequent decades expanded the dimensionality of the problem by incorporating photometric information, particularly colour--magnitude diagram constraints, as well as improved spatial modelling. These developments enabled more robust separation of cluster and field populations, although most approaches still relied on simplifying assumptions such as Gaussian distributions and separability of observables.

The advent of the \textit{Gaia} mission represents a paradigm shift in the field. The availability of high-precision astrometry and photometry in multi-dimensional space has enabled the application of non-parametric, clustering-based, and machine-learning methods. Algorithms such as Gaussian Mixture Models, UPMASK, DBSCAN, and HDBSCAN have significantly improved membership determinations, particularly in complex or sparse stellar systems. In parallel, large homogeneous catalogues of cluster members have been constructed, providing a uniform basis for Galactic studies.

Despite these advances, several fundamental challenges remain. Field-star populations exhibit complex, non-Gaussian structures, unresolved binaries introduce systematic biases, and extinction variations complicate photometric interpretations. Moreover, machine-learning approaches, while powerful, often lack physical interpretability and depend strongly on training data quality.

Across all methodologies, a key conclusion is that cluster membership determination remains an intrinsically ill-posed inverse problem. No single method is universally optimal; instead, performance depends strongly on data dimensionality, observational uncertainties, and the dynamical state of the cluster.

Future progress will likely be driven by the combination of next-generation surveys, such as LSST and the Roman Space Telescope, with advanced statistical and machine-learning techniques. In particular, probabilistic deep learning, graph-based models, and physics-informed approaches are expected to play a central role in future developments. These methods will enable the transition from static classification to fully dynamical, time-dependent inference of stellar populations.

In summary, the field has evolved from simple spatial classification to a highly sophisticated, multi-dimensional statistical discipline. Continued progress will depend on the integration of physical modelling, statistical rigour, and scalable computational methods capable of handling the complexity of forthcoming astronomical datasets.

\begin{acknowledgements}

This work was supported by the grant GA{\v C}R 23-07605S.

\end{acknowledgements}

\bibliographystyle{aa} 
\bibliography{paper.bib}

@ARTICLE{Trumpler1930,
       author = {{Trumpler}, Robert Julius},
        title = "{Preliminary results on the distances, dimensions and space distribution of open star clusters}",
      journal = {Lick Observatory Bulletin},
         year = 1930,
        month = jan,
       volume = {420},
        pages = {154-188},
          doi = {10.5479/ADS/bib/1930LicOB.14.154T},
       adsurl = {https://ui.adsabs.harvard.edu/abs/1930LicOB..14..154T}
}

@ARTICLE{sanders1971,
       author = {{Sanders}, W.~L.},
        title = "{An improved method for computing membership probabilities in open clusters.}",
      journal = {\aap},
         year = 1971,
        month = sep,
       volume = {14},
        pages = {226-232},
       adsurl = {https://ui.adsabs.harvard.edu/abs/1971A&A....14..226S}
}

@ARTICLE{stetson1987,
       author = {{Stetson}, Peter B.},
        title = "{DAOPHOT: A Computer Program for Crowded-Field Stellar Photometry}",
      journal = {\pasp},
         year = 1987,
        month = mar,
       volume = {99},
        pages = {191},
          doi = {10.1086/131977},
       adsurl = {https://ui.adsabs.harvard.edu/abs/1987PASP...99..191S}
}

@ARTICLE{zhao1990,
       author = {{Zhao}, J.~L. and {He}, Y.~P.},
        title = "{An improved method for membership determination of stellar clusters with proper motions with different accuracies.}",
      journal = {\aap},
         year = 1990,
        month = oct,
       volume = {237},
        pages = {54},
       adsurl = {https://ui.adsabs.harvard.edu/abs/1990A&A...237...54Z}
}

@ARTICLE{gaia2016,
       author = {{Gaia Collaboration} and {Prusti}, T. and {de Bruijne}, J.~H.~J. and {Brown}, A.~G.~A. and {Vallenari}, A. and {Babusiaux}, C. and {Bailer-Jones}, C.~A.~L. and {Bastian}, U. and {Biermann}, M. and {Evans}, D.~W. and {Eyer}, L. and {Jansen}, F. and {Jordi}, C. and {Klioner}, S.~A. and {Lammers}, U. and {Lindegren}, L. and {Luri}, X. and {Mignard}, F. and {Milligan}, D.~J. and {Panem}, C. and {Poinsignon}, V. and {Pourbaix}, D. and {Randich}, S. and {Sarri}, G. and {Sartoretti}, P. and {Siddiqui}, H.~I. and {Soubiran}, C. and {Valette}, V. and {van Leeuwen}, F. and {Walton}, N.~A. and {Aerts}, C. and {Arenou}, F. and {Cropper}, M. and {Drimmel}, R. and {H{\o}g}, E. and {Katz}, D. and {Lattanzi}, M.~G. and {O'Mullane}, W. and {Grebel}, E.~K. and {Holland}, A.~D. and {Huc}, C. and {Passot}, X. and {Bramante}, L. and {Cacciari}, C. and {Casta{\~n}eda}, J. and {Chaoul}, L. and {Cheek}, N. and {De Angeli}, F. and {Fabricius}, C. and {Guerra}, R. and {Hern{\'a}ndez}, J. and {Jean-Antoine-Piccolo}, A. and {Masana}, E. and {Messineo}, R. and {Mowlavi}, N. and {Nienartowicz}, K. and {Ord{\'o}{\~n}ez-Blanco}, D. and {Panuzzo}, P. and {Portell}, J. and {Richards}, P.~J. and {Riello}, M. and {Seabroke}, G.~M. and {Tanga}, P. and {Th{\'e}venin}, F. and {Torra}, J. and {Els}, S.~G. and {Gracia-Abril}, G. and {Comoretto}, G. and {Garcia-Reinaldos}, M. and {Lock}, T. and {Mercier}, E. and {Altmann}, M. and {Andrae}, R. and {Astraatmadja}, T.~L. and {Bellas-Velidis}, I. and {Benson}, K. and {Berthier}, J. and {Blomme}, R. and {Busso}, G. and {Carry}, B. and {Cellino}, A. and {Clementini}, G. and {Cowell}, S. and {Creevey}, O. and {Cuypers}, J. and {Davidson}, M. and {De Ridder}, J. and {de Torres}, A. and {Delchambre}, L. and {Dell'Oro}, A. and {Ducourant}, C. and {Fr{\'e}mat}, Y. and {Garc{\'\i}a-Torres}, M. and {Gosset}, E. and {Halbwachs}, J.-L. and {Hambly}, N.~C. and {Harrison}, D.~L. and {Hauser}, M. and {Hestroffer}, D. and {Hodgkin}, S.~T. and {Huckle}, H.~E. and {Hutton}, A. and {Jasniewicz}, G. and {Jordan}, S. and {Kontizas}, M. and {Korn}, A.~J. and {Lanzafame}, A.~C. and {Manteiga}, M. and {Moitinho}, A. and {Muinonen}, K. and {Osinde}, J. and {Pancino}, E. and {Pauwels}, T. and {Petit}, J.-M. and {Recio-Blanco}, A. and {Robin}, A.~C. and {Sarro}, L.~M. and {Siopis}, C. and {Smith}, M. and {Smith}, K.~W. and {Sozzetti}, A. and {Thuillot}, W. and {van Reeven}, W. and {Viala}, Y. and {Abbas}, U. and {Abreu Aramburu}, A. and {Accart}, S. and {Aguado}, J.~J. and {Allan}, P.~M. and {Allasia}, W. and {Altavilla}, G. and {{\'A}lvarez}, M.~A. and {Alves}, J. and {Anderson}, R.~I. and {Andrei}, A.~H. and {Anglada Varela}, E. and {Antiche}, E. and {Antoja}, T. and {Ant{\'o}n}, S. and {Arcay}, B. and {Atzei}, A. and {Ayache}, L. and {Bach}, N. and {Baker}, S.~G. and {Balaguer-N{\'u}{\~n}ez}, L. and {Barache}, C. and {Barata}, C. and {Barbier}, A. and {Barblan}, F. and {Baroni}, M. and {Barrado y Navascu{\'e}s}, D. and {Barros}, M. and {Barstow}, M.~A. and {Becciani}, U. and {Bellazzini}, M. and {Bellei}, G. and {Bello Garc{\'\i}a}, A. and {Belokurov}, V. and {Bendjoya}, P. and {Berihuete}, A. and {Bianchi}, L. and {Bienaym{\'e}}, O. and {Billebaud}, F. and {Blagorodnova}, N. and {Blanco-Cuaresma}, S. and {Boch}, T. and {Bombrun}, A. and {Borrachero}, R. and {Bouquillon}, S. and {Bourda}, G. and {Bouy}, H. and {Bragaglia}, A. and {Breddels}, M.~A. and {Brouillet}, N. and {Br{\"u}semeister}, T. and {Bucciarelli}, B. and {Budnik}, F. and {Burgess}, P. and {Burgon}, R. and {Burlacu}, A. and {Busonero}, D. and {Buzzi}, R. and {Caffau}, E. and {Cambras}, J. and {Campbell}, H. and {Cancelliere}, R. and {Cantat-Gaudin}, T. and {Carlucci}, T. and {Carrasco}, J.~M. and {Castellani}, M. and {Charlot}, P. and {Charnas}, J. and {Charvet}, P. and {Chassat}, F. and {Chiavassa}, A. and {Clotet}, M. and {Cocozza}, G. and {Collins}, R.~S. and {Collins}, P. and {Costigan}, G.},
        title = "{The Gaia mission}",
      journal = {\aap},
         year = 2016,
        month = nov,
       volume = {595},
          eid = {A1},
        pages = {A1},
          doi = {10.1051/0004-6361/201629272},
archivePrefix = {arXiv},
       eprint = {1609.04153},
 primaryClass = {astro-ph.IM},
       adsurl = {https://ui.adsabs.harvard.edu/abs/2016A&A...595A...1G}
}

@ARTICLE{bonatto2007,
       author = {{Bonatto}, C. and {Bica}, E.},
        title = "{Open clusters in dense fields: the importance of field-star decontamination for NGC 5715, Lyng{\r{a}} 4, Lyng{\r{a}} 9, Trumpler 23, Trumpler 26 and Czernik 37}",
      journal = {\mnras},
         year = 2007,
        month = may,
       volume = {377},
       number = {3},
        pages = {1301-1323},
          doi = {10.1111/j.1365-2966.2007.11691.x},
archivePrefix = {arXiv},
       eprint = {astro-ph/0703025},
 primaryClass = {astro-ph},
       adsurl = {https://ui.adsabs.harvard.edu/abs/2007MNRAS.377.1301B}
}

@ARTICLE{gaia2018dr2,
       author = {{Gaia Collaboration} and {Brown}, A.~G.~A. and {Vallenari}, A. and {Prusti}, T. and {de Bruijne}, J.~H.~J. and {Babusiaux}, C. and {Bailer-Jones}, C.~A.~L. and {Biermann}, M. and {Evans}, D.~W. and {Eyer}, L. and {Jansen}, F. and {Jordi}, C. and {Klioner}, S.~A. and {Lammers}, U. and {Lindegren}, L. and {Luri}, X. and {Mignard}, F. and {Panem}, C. and {Pourbaix}, D. and {Randich}, S. and {Sartoretti}, P. and {Siddiqui}, H.~I. and {Soubiran}, C. and {van Leeuwen}, F. and {Walton}, N.~A. and {Arenou}, F. and {Bastian}, U. and {Cropper}, M. and {Drimmel}, R. and {Katz}, D. and {Lattanzi}, M.~G. and {Bakker}, J. and {Cacciari}, C. and {Casta{\~n}eda}, J. and {Chaoul}, L. and {Cheek}, N. and {De Angeli}, F. and {Fabricius}, C. and {Guerra}, R. and {Holl}, B. and {Masana}, E. and {Messineo}, R. and {Mowlavi}, N. and {Nienartowicz}, K. and {Panuzzo}, P. and {Portell}, J. and {Riello}, M. and {Seabroke}, G.~M. and {Tanga}, P. and {Th{\'e}venin}, F. and {Gracia-Abril}, G. and {Comoretto}, G. and {Garcia-Reinaldos}, M. and {Teyssier}, D. and {Altmann}, M. and {Andrae}, R. and {Audard}, M. and {Bellas-Velidis}, I. and {Benson}, K. and {Berthier}, J. and {Blomme}, R. and {Burgess}, P. and {Busso}, G. and {Carry}, B. and {Cellino}, A. and {Clementini}, G. and {Clotet}, M. and {Creevey}, O. and {Davidson}, M. and {De Ridder}, J. and {Delchambre}, L. and {Dell'Oro}, A. and {Ducourant}, C. and {Fern{\'a}ndez-Hern{\'a}ndez}, J. and {Fouesneau}, M. and {Fr{\'e}mat}, Y. and {Galluccio}, L. and {Garc{\'\i}a-Torres}, M. and {Gonz{\'a}lez-N{\'u}{\~n}ez}, J. and {Gonz{\'a}lez-Vidal}, J.~J. and {Gosset}, E. and {Guy}, L.~P. and {Halbwachs}, J.-L. and {Hambly}, N.~C. and {Harrison}, D.~L. and {Hern{\'a}ndez}, J. and {Hestroffer}, D. and {Hodgkin}, S.~T. and {Hutton}, A. and {Jasniewicz}, G. and {Jean-Antoine-Piccolo}, A. and {Jordan}, S. and {Korn}, A.~J. and {Krone-Martins}, A. and {Lanzafame}, A.~C. and {Lebzelter}, T. and {L{\"o}ffler}, W. and {Manteiga}, M. and {Marrese}, P.~M. and {Mart{\'\i}n-Fleitas}, J.~M. and {Moitinho}, A. and {Mora}, A. and {Muinonen}, K. and {Osinde}, J. and {Pancino}, E. and {Pauwels}, T. and {Petit}, J.-M. and {Recio-Blanco}, A. and {Richards}, P.~J. and {Rimoldini}, L. and {Robin}, A.~C. and {Sarro}, L.~M. and {Siopis}, C. and {Smith}, M. and {Sozzetti}, A. and {S{\"u}veges}, M. and {Torra}, J. and {van Reeven}, W. and {Abbas}, U. and {Abreu Aramburu}, A. and {Accart}, S. and {Aerts}, C. and {Altavilla}, G. and {{\'A}lvarez}, M.~A. and {Alvarez}, R. and {Alves}, J. and {Anderson}, R.~I. and {Andrei}, A.~H. and {Anglada Varela}, E. and {Antiche}, E. and {Antoja}, T. and {Arcay}, B. and {Astraatmadja}, T.~L. and {Bach}, N. and {Baker}, S.~G. and {Balaguer-N{\'u}{\~n}ez}, L. and {Balm}, P. and {Barache}, C. and {Barata}, C. and {Barbato}, D. and {Barblan}, F. and {Barklem}, P.~S. and {Barrado}, D. and {Barros}, M. and {Barstow}, M.~A. and {Bartholom{\'e} Mu{\~n}oz}, S. and {Bassilana}, J.-L. and {Becciani}, U. and {Bellazzini}, M. and {Berihuete}, A. and {Bertone}, S. and {Bianchi}, L. and {Bienaym{\'e}}, O. and {Blanco-Cuaresma}, S. and {Boch}, T. and {Boeche}, C. and {Bombrun}, A. and {Borrachero}, R. and {Bossini}, D. and {Bouquillon}, S. and {Bourda}, G. and {Bragaglia}, A. and {Bramante}, L. and {Breddels}, M.~A. and {Bressan}, A. and {Brouillet}, N. and {Br{\"u}semeister}, T. and {Brugaletta}, E. and {Bucciarelli}, B. and {Burlacu}, A. and {Busonero}, D. and {Butkevich}, A.~G. and {Buzzi}, R. and {Caffau}, E. and {Cancelliere}, R. and {Cannizzaro}, G. and {Cantat-Gaudin}, T. and {Carballo}, R. and {Carlucci}, T. and {Carrasco}, J.~M. and {Casamiquela}, L. and {Castellani}, M. and {Castro-Ginard}, A. and {Charlot}, P. and {Chemin}, L. and {Chiavassa}, A. and {Cocozza}, G. and {Costigan}, G. and {Cowell}, S. and {Crifo}, F. and {Crosta}, M. and {Crowley}, C. and {Cuypers}, J. and {Dafonte}, C. and {Damerdji}, Y. and {Dapergolas}, A. and {David}, P. and {David}, M. and {de Laverny}, P. and {De Luise}, F.},
        title = "{Gaia Data Release 2. Summary of the contents and survey properties}",
      journal = {\aap},
         year = 2018,
        month = aug,
       volume = {616},
          eid = {A1},
        pages = {A1},
          doi = {10.1051/0004-6361/201833051},
archivePrefix = {arXiv},
       eprint = {1804.09365},
 primaryClass = {astro-ph.GA},
       adsurl = {https://ui.adsabs.harvard.edu/abs/2018A&A...616A...1G}
}

@ARTICLE{gaia2022dr3,
       author = {{Gaia Collaboration} and {Vallenari}, A. and {Brown}, A.~G.~A. and {Prusti}, T. and {de Bruijne}, J.~H.~J. and {Arenou}, F. and {Babusiaux}, C. and {Biermann}, M. and {Creevey}, O.~L. and {Ducourant}, C. and {Evans}, D.~W. and {Eyer}, L. and {Guerra}, R. and {Hutton}, A. and {Jordi}, C. and {Klioner}, S.~A. and {Lammers}, U.~L. and {Lindegren}, L. and {Luri}, X. and {Mignard}, F. and {Panem}, C. and {Pourbaix}, D. and {Randich}, S. and {Sartoretti}, P. and {Soubiran}, C. and {Tanga}, P. and {Walton}, N.~A. and {Bailer-Jones}, C.~A.~L. and {Bastian}, U. and {Drimmel}, R. and {Jansen}, F. and {Katz}, D. and {Lattanzi}, M.~G. and {van Leeuwen}, F. and {Bakker}, J. and {Cacciari}, C. and {Casta{\~n}eda}, J. and {De Angeli}, F. and {Fabricius}, C. and {Fouesneau}, M. and {Fr{\'e}mat}, Y. and {Galluccio}, L. and {Guerrier}, A. and {Heiter}, U. and {Masana}, E. and {Messineo}, R. and {Mowlavi}, N. and {Nicolas}, C. and {Nienartowicz}, K. and {Pailler}, F. and {Panuzzo}, P. and {Riclet}, F. and {Roux}, W. and {Seabroke}, G.~M. and {Sordo}, R. and {Th{\'e}venin}, F. and {Gracia-Abril}, G. and {Portell}, J. and {Teyssier}, D. and {Altmann}, M. and {Andrae}, R. and {Audard}, M. and {Bellas-Velidis}, I. and {Benson}, K. and {Berthier}, J. and {Blomme}, R. and {Burgess}, P.~W. and {Busonero}, D. and {Busso}, G. and {C{\'a}novas}, H. and {Carry}, B. and {Cellino}, A. and {Cheek}, N. and {Clementini}, G. and {Damerdji}, Y. and {Davidson}, M. and {de Teodoro}, P. and {Nu{\~n}ez Campos}, M. and {Delchambre}, L. and {Dell'Oro}, A. and {Esquej}, P. and {Fern{\'a}ndez-Hern{\'a}ndez}, J. and {Fraile}, E. and {Garabato}, D. and {Garc{\'\i}a-Lario}, P. and {Gosset}, E. and {Haigron}, R. and {Halbwachs}, J.-L. and {Hambly}, N.~C. and {Harrison}, D.~L. and {Hern{\'a}ndez}, J. and {Hestroffer}, D. and {Hodgkin}, S.~T. and {Holl}, B. and {Jan{\ss}en}, K. and {Jevardat de Fombelle}, G. and {Jordan}, S. and {Krone-Martins}, A. and {Lanzafame}, A.~C. and {L{\"o}ffler}, W. and {Marchal}, O. and {Marrese}, P.~M. and {Moitinho}, A. and {Muinonen}, K. and {Osborne}, P. and {Pancino}, E. and {Pauwels}, T. and {Recio-Blanco}, A. and {Reyl{\'e}}, C. and {Riello}, M. and {Rimoldini}, L. and {Roegiers}, T. and {Rybizki}, J. and {Sarro}, L.~M. and {Siopis}, C. and {Smith}, M. and {Sozzetti}, A. and {Utrilla}, E. and {van Leeuwen}, M. and {Abbas}, U. and {{\'A}brah{\'a}m}, P. and {Abreu Aramburu}, A. and {Aerts}, C. and {Aguado}, J.~J. and {Ajaj}, M. and {Aldea-Montero}, F. and {Altavilla}, G. and {{\'A}lvarez}, M.~A. and {Alves}, J. and {Anders}, F. and {Anderson}, R.~I. and {Anglada Varela}, E. and {Antoja}, T. and {Baines}, D. and {Baker}, S.~G. and {Balaguer-N{\'u}{\~n}ez}, L. and {Balbinot}, E. and {Balog}, Z. and {Barache}, C. and {Barbato}, D. and {Barros}, M. and {Barstow}, M.~A. and {Bartolom{\'e}}, S. and {Bassilana}, J.-L. and {Bauchet}, N. and {Becciani}, U. and {Bellazzini}, M. and {Berihuete}, A. and {Bernet}, M. and {Bertone}, S. and {Bianchi}, L. and {Binnenfeld}, A. and {Blanco-Cuaresma}, S. and {Blazere}, A. and {Boch}, T. and {Bombrun}, A. and {Bossini}, D. and {Bouquillon}, S. and {Bragaglia}, A. and {Bramante}, L. and {Breedt}, E. and {Bressan}, A. and {Brouillet}, N. and {Brugaletta}, E. and {Bucciarelli}, B. and {Burlacu}, A. and {Butkevich}, A.~G. and {Buzzi}, R. and {Caffau}, E. and {Cancelliere}, R. and {Cantat-Gaudin}, T. and {Carballo}, R. and {Carlucci}, T. and {Carnerero}, M.~I. and {Carrasco}, J.~M. and {Casamiquela}, L. and {Castellani}, M. and {Castro-Ginard}, A. and {Chaoul}, L. and {Charlot}, P. and {Chemin}, L. and {Chiaramida}, V. and {Chiavassa}, A. and {Chornay}, N. and {Comoretto}, G. and {Contursi}, G. and {Cooper}, W.~J. and {Cornez}, T. and {Cowell}, S. and {Crifo}, F. and {Cropper}, M. and {Crosta}, M. and {Crowley}, C. and {Dafonte}, C. and {Dapergolas}, A. and {David}, M. and {David}, P. and {de Laverny}, P. and {De Luise}, F. and {De March}, R.},
        title = "{Gaia Data Release 3. Summary of the content and survey properties}",
      journal = {\aap},
         year = 2023,
        month = jun,
       volume = {674},
          eid = {A1},
        pages = {A1},
          doi = {10.1051/0004-6361/202243940},
archivePrefix = {arXiv},
       eprint = {2208.00211},
 primaryClass = {astro-ph.GA},
       adsurl = {https://ui.adsabs.harvard.edu/abs/2023A&A...674A...1G}
}

@ARTICLE{cantatgaudin2018,
       author = {{Cantat-Gaudin}, T. and {Jordi}, C. and {Vallenari}, A. and {Bragaglia}, A. and {Balaguer-N{\'u}{\~n}ez}, L. and {Soubiran}, C. and {Bossini}, D. and {Moitinho}, A. and {Castro-Ginard}, A. and {Krone-Martins}, A. and {Casamiquela}, L. and {Sordo}, R. and {Carrera}, R.},
        title = "{A Gaia DR2 view of the open cluster population in the Milky Way}",
      journal = {\aap},
         year = 2018,
        month = oct,
       volume = {618},
          eid = {A93},
        pages = {A93},
          doi = {10.1051/0004-6361/201833476},
archivePrefix = {arXiv},
       eprint = {1805.08726},
 primaryClass = {astro-ph.GA},
       adsurl = {https://ui.adsabs.harvard.edu/abs/2018A&A...618A..93C}
}

@ARTICLE{balaguer2004,
       author = {{Balaguer-N{\'u}{\~n}ez}, L. and {Jordi}, C. and {Galad{\'\i}-Enr{\'\i}quez}, D. and {Zhao}, J.~L.},
        title = "{New membership determination and proper motions of NGC 1817. Parametric and non-parametric approach}",
      journal = {\aap},
         year = 2004,
        month = nov,
       volume = {426},
        pages = {819-826},
          doi = {10.1051/0004-6361:20041332},
archivePrefix = {arXiv},
       eprint = {astro-ph/0407455},
 primaryClass = {astro-ph},
       adsurl = {https://ui.adsabs.harvard.edu/abs/2004A&A...426..819B}
}

@ARTICLE{cantatgaudin2020,
       author = {{Cantat-Gaudin}, T. and {Anders}, F.},
        title = "{Clusters and mirages: cataloguing stellar aggregates in the Milky Way}",
      journal = {\aap},
         year = 2020,
        month = jan,
       volume = {633},
          eid = {A99},
        pages = {A99},
          doi = {10.1051/0004-6361/201936691},
archivePrefix = {arXiv},
       eprint = {1911.07075},
 primaryClass = {astro-ph.SR},
       adsurl = {https://ui.adsabs.harvard.edu/abs/2020A&A...633A..99C}
}

@ARTICLE{castroginard2020,
       author = {{Castro-Ginard}, A. and {Jordi}, C. and {Luri}, X. and {{\'A}lvarez Cid-Fuentes}, J. and {Casamiquela}, L. and {Anders}, F. and {Cantat-Gaudin}, T. and {Mongui{\'o}}, M. and {Balaguer-N{\'u}{\~n}ez}, L. and {Sol{\`a}}, S. and {Badia}, R.~M.},
        title = "{Hunting for open clusters in Gaia DR2: 582 new open clusters in the Galactic disc}",
      journal = {\aap},
         year = 2020,
        month = mar,
       volume = {635},
          eid = {A45},
        pages = {A45},
          doi = {10.1051/0004-6361/201937386},
archivePrefix = {arXiv},
       eprint = {2001.07122},
 primaryClass = {astro-ph.GA},
       adsurl = {https://ui.adsabs.harvard.edu/abs/2020A&A...635A..45C}
}

@ARTICLE{dias2021,
       author = {{Dias}, W.~S. and {Monteiro}, H. and {Moitinho}, A. and {L{\'e}pine}, J.~R.~D. and {Carraro}, G. and {Paunzen}, E. and {Alessi}, B. and {Villela}, L.},
        title = "{Updated parameters of 1743 open clusters based on Gaia DR2}",
      journal = {\mnras},
         year = 2021,
        month = jun,
       volume = {504},
       number = {1},
        pages = {356-371},
          doi = {10.1093/mnras/stab770},
archivePrefix = {arXiv},
       eprint = {2103.12829},
 primaryClass = {astro-ph.SR},
       adsurl = {https://ui.adsabs.harvard.edu/abs/2021MNRAS.504..356D}
}

@ARTICLE{kharchenko2004,
       author = {{Kharchenko}, N.~V. and {Piskunov}, A.~E. and {R{\"o}ser}, S. and {Schilbach}, E. and {Scholz}, R.-D.},
        title = "{Astrophysical supplements to the ASCC-2.5. II. Membership probabilities in 520 Galactic open cluster sky areas}",
      journal = {Astronomische Nachrichten},
         year = 2004,
        month = dec,
       volume = {325},
       number = {9},
        pages = {740-748},
          doi = {10.1002/asna.200410256},
       adsurl = {https://ui.adsabs.harvard.edu/abs/2004AN....325..740K}
}

@ARTICLE{kharchenko2013,
       author = {{Kharchenko}, N.~V. and {Piskunov}, A.~E. and {Schilbach}, E. and {R{\"o}ser}, S. and {Scholz}, R.-D.},
        title = "{Global survey of star clusters in the Milky Way. II. The catalogue of basic parameters}",
      journal = {\aap},
         year = 2013,
        month = oct,
       volume = {558},
          eid = {A53},
        pages = {A53},
          doi = {10.1051/0004-6361/201322302},
archivePrefix = {arXiv},
       eprint = {1308.5822},
 primaryClass = {astro-ph.GA},
       adsurl = {https://ui.adsabs.harvard.edu/abs/2013A&A...558A..53K}
}

@ARTICLE{krone2014,
       author = {{Krone-Martins}, A. and {Moitinho}, A.},
        title = "{UPMASK: unsupervised photometric membership assignment in stellar clusters}",
      journal = {\aap},
         year = 2014,
        month = jan,
       volume = {561},
          eid = {A57},
        pages = {A57},
          doi = {10.1051/0004-6361/201321143},
archivePrefix = {arXiv},
       eprint = {1309.4471},
 primaryClass = {astro-ph.IM},
       adsurl = {https://ui.adsabs.harvard.edu/abs/2014A&A...561A..57K}
}

@ARTICLE{portegies2010,
       author = {{Portegies Zwart}, Simon F. and {McMillan}, Stephen L.~W. and {Gieles}, Mark},
        title = "{Young Massive Star Clusters}",
      journal = {\araa},
         year = 2010,
        month = sep,
       volume = {48},
        pages = {431-493},
          doi = {10.1146/annurev-astro-081309-130834},
archivePrefix = {arXiv},
       eprint = {1002.1961},
 primaryClass = {astro-ph.GA},
       adsurl = {https://ui.adsabs.harvard.edu/abs/2010ARA&A..48..431P}
}

@dataset{lynga1987,
       author = {{Lynga}, G.},
        title = "{VizieR Online Data Catalog: Open Cluster Data 5th Edition (Lynga 1987)}",
 howpublished = {VizieR On-line Data Catalog: VII/92A.  Originally published in: Lund Observatory},
         year = 1995,
        month = feb,
          eid = {VII/92A},
       adsurl = {https://ui.adsabs.harvard.edu/abs/1995yCat.7092....0L}
}

@ARTICLE{janes1988,
       author = {{Janes}, K. and {Adler}, D.},
        title = "{Open clusters and galactic structure.}",
      journal = {\apjs},
         year = 1982,
        month = jul,
       volume = {49},
        pages = {425-446},
          doi = {10.1086/190805},
       adsurl = {https://ui.adsabs.harvard.edu/abs/1982ApJS...49..425J}
}

@ARTICLE{vasilevskis1958,
       author = {{Vasilevskis}, S. and {Klemola}, A. and {Preston}, G.},
        title = "{Relative proper motions of stars in the region of the open cluster NGC 6633.}",
      journal = {\aj},
         year = 1958,
        month = jan,
       volume = {63},
        pages = {387-395},
          doi = {10.1086/107787},
       adsurl = {https://ui.adsabs.harvard.edu/abs/1958AJ.....63..387V}
}

@ARTICLE{brosche1989,
       author = {{Brosche}, P. and {Wildermann}, E. and {Geffert}, M.},
        title = "{Astrometric plate reductions with orthogonal functions}",
      journal = {\aap},
         year = 1989,
        month = feb,
       volume = {211},
       number = {1},
        pages = {239-244},
       adsurl = {https://ui.adsabs.harvard.edu/abs/1989A&A...211..239B}
}

@ARTICLE{2023A&A...673A.114H,
       author = {{Hunt}, Emily L. and {Reffert}, Sabine},
        title = "{Improving the open cluster census. II. An all-sky cluster catalogue with Gaia DR3}",
      journal = {\aap},
         year = 2023,
        month = may,
       volume = {673},
          eid = {A114},
        pages = {A114},
          doi = {10.1051/0004-6361/202346285},
archivePrefix = {arXiv},
       eprint = {2303.13424},
 primaryClass = {astro-ph.GA},
       adsurl = {https://ui.adsabs.harvard.edu/abs/2023A&A...673A.114H}
}

@ARTICLE{2023MNRAS.526.4107P,
       author = {{Perren}, Gabriel I. and {Pera}, Mar{\'\i}a S. and {Navone}, Hugo D. and {V{\'a}zquez}, Rub{\'e}n A.},
        title = "{The Unified Cluster Catalogue: towards a comprehensive and homogeneous data base of stellar clusters}",
      journal = {\mnras},
         year = 2023,
        month = dec,
       volume = {526},
       number = {3},
        pages = {4107-4119},
          doi = {10.1093/mnras/stad2826},
       adsurl = {https://ui.adsabs.harvard.edu/abs/2023MNRAS.526.4107P}
}

\begin{appendix} 

\section{Tables}
\label{Appendix A}

\begin{table*}
\centering
\caption{Early approaches to cluster membership determination (1930--1957).}
\label{tab:early_methods}
\begin{tabular}{p{3.5cm} p{4.5cm} p{4.5cm}}
\hline
Method & Principle & Limitations \\
\hline
Radial star counts & Overdensity relative to background field & Severe contamination in dense Galactic fields \\
Visual inspection & Identification of concentrated stellar groups on plates & Highly subjective, non-reproducible \\
Control field subtraction & Statistical subtraction of background density & Assumes uniform field distribution \\
Early photographic astrometry & Identification of common motion over long baselines & Large uncertainties, small sample sizes \\
\hline
\end{tabular}
\end{table*}

\begin{table*}
\centering
\caption{First photographic investigations in cluster membership studies (1930--1957).}
\label{tab:photo_era}
\begin{tabular}{p{4cm} p{5cm} p{5cm}}
\hline
Approach & Principle & Main Limitations \\
\hline
Single photographic plate analysis & Identification of overdensities in star fields & No kinematic information; strong field contamination \\
Multi-epoch plate comparison & Detection of relative positional shifts over time & Large systematic errors; short baselines \\
Magnitude-limited samples & Restriction to bright stars for cleaner samples & Severe incompleteness; biased toward massive stars \\
Visual plate blinking & Qualitative identification of moving stars & Subjective, non-reproducible results \\
\hline
\end{tabular}
\end{table*}

\begin{table*}
\centering
\caption{Transition from spatial to kinematic membership determination (1958--1965).}
\label{tab:pm_advantages}
\begin{tabular}{p{4cm} p{5.5cm} p{5.5cm}}
\hline
Method & Improvement & Remaining limitations \\
\hline
Spatial star counts & Simple implementation & Severe field contamination \\
Photographic astrometry & Introduction of motion information & Large measurement errors \\
Bivariate kinematic separation & First statistical population model & Simplified Gaussian assumptions \\
Early probabilistic interpretation & Membership likelihood concept introduced & Not yet fully Bayesian or rigorous \\
\hline
\end{tabular}
\end{table*}

\begin{table*}
\centering
\caption{Evolution toward statistical likelihood-based membership determination (1965--1970).}
\label{tab:transition_ml}
\begin{tabular}{p{4cm} p{5.5cm} p{5.5cm}}
\hline
Approach & Key Advancement & Remaining limitations \\
\hline
Gaussian cluster modeling & Introduction of parametric population models & Field population poorly constrained \\
Anisotropic field distributions & Improved realism for Galactic kinematics & Still empirically tuned \\
Likelihood-based classification & First formal probability framework & Not yet fully Bayesian or error-consistent \\
Photographic proper-motion refinement & Improved astrometric precision & Limited sample sizes \\
\hline
\end{tabular}
\end{table*}

\begin{table*}
\centering
\caption{Classical maximum-likelihood membership methods (1971--1990).}
\label{tab:ml_1970_1990}
\begin{tabular}{p{4cm} p{5.5cm} p{5.5cm}}
\hline
Method & Key Idea & Limitation \\
\hline
\citet{sanders1971} model & Gaussian mixture in proper-motion space & Simplified field distribution \\
Iterative parameter estimation & Joint estimation of cluster + field parameters & Sensitive to initial conditions \\
Improved astrometric reductions & Reduced observational noise & Still plate-based systematics \\
Extended Gaussian field models & More realistic Galactic kinematics & Limited flexibility for complex fields \\
\hline
\end{tabular}
\end{table*}

\clearpage

\begin{table*}
\centering
\caption{Cluster membership methods in the photometric--astrometric era (1990--2010).}
\label{tab:ccd_era_methods}
\begin{tabular}{p{4cm} p{5.5cm} p{5.5cm}}
\hline
Method & Key Idea & Limitation \\
\hline
CMD filtering & Isochrone-based selection in color--magnitude space & Sensitive to reddening and binaries \\
Spatial + photometric combination & Joint use of radial profiles and CMD position & Often assumes independence of variables \\
Kernel density estimation (KDE) & Non-parametric modeling of distributions & Bandwidth sensitivity \\
Field decontamination & Statistical subtraction of field CMD & Requires representative control field \\
Early multi-dimensional models & Combined astrometric + photometric criteria & No unified probabilistic framework \\
\hline
\end{tabular}
\end{table*}

\begin{table*}
\centering
\caption{Cluster membership methods in the $Gaia$ era (2016--present).}
\label{tab:gaia_methods}
\begin{tabular}{p{4cm} p{5.5cm} p{5.5cm}}
\hline
Method & Key Idea & Limitation \\
\hline
Gaussian Mixture Models (GMM) & Probabilistic mixture in high-dimensional space & Assumes Gaussian components \\
Bayesian inference & Prior-informed probabilistic modeling & Sensitive to priors \\
UPMASK & Unsupervised clustering + spatial validation & Computationally expensive \\
DBSCAN/HDBSCAN & Density-based clustering in phase space & Parameter tuning sensitivity \\
Random Forest / ML models & Supervised classification from training sets & Training bias dependence \\
Cantat-Gaudin pipeline & Combined astrometric + photometric filtering & Limited to $Gaia$ completeness regime \\
\hline
\end{tabular}
\end{table*}

\begin{table*}
\centering
\caption{Comparison of major families of cluster membership determination methods.}
\label{tab:method_families}
\begin{tabular}{p{3.5cm} p{3.5cm} p{4cm} p{2.5cm} p{3.5cm}}
\hline
Method family & Input space & Statistical assumption & Strengths & Limitations \\
\hline
Spatial methods & (RA, Dec) & None or uniform field model & Simple, early applicability & Severe contamination, no kinematics \\
Kinematic (classical) & ($\mu_\alpha, \mu_\delta$) & Gaussian cluster + empirical field & First probabilistic models & Strong Gaussian assumptions \\
Photometric methods & (CMD space) & Isochrone-based distribution & Effective for young clusters & Sensitive to reddening and binaries \\
Maximum likelihood / Bayesian & Multi-dimensional & Parametric likelihood models & Statistically consistent framework & Model dependence, computational cost \\
Machine learning & High-dimensional feature space & Non-parametric / learned decision boundary & Handles complex structures & Requires training data, less interpretable \\
Clustering algorithms (DBSCAN/HDBSCAN) & Phase space density & Density-based assumption & No prior model required & Parameter tuning sensitivity \\
\hline
\end{tabular}
\end{table*}

\clearpage

\section{Figures}
\label{Appendix B}

\begin{figure*}
\centering
\begin{tikzpicture}[scale=1.0]

\draw[->] (0,0,0) -- (4,0,0) node[right] {$\mu_\alpha$};
\draw[->] (0,0,0) -- (0,4,0) node[above] {$\mu_\delta$};
\draw[->] (0,0,0) -- (0,0,4) node[below] {Color};

\draw[blue, thick] (2,2,2) circle (0.6);
\node[blue] at (3.3,2.7,2) {Cluster locus};

\draw[red, thick] (1,3,1) ellipse (1.2 and 0.6);
\node[red] at (1.8,3.8,1) {Field population};

\end{tikzpicture}

\caption{Conceptual representation of multi-dimensional membership space combining astrometric and photometric information. The cluster forms a compact locus, while field stars are broadly distributed.}
\label{fig:multidim_space}
\end{figure*}
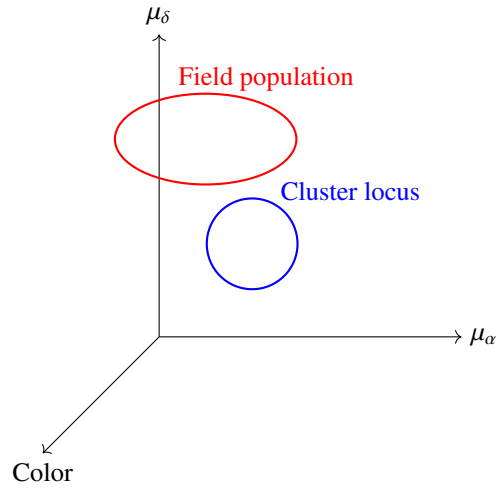

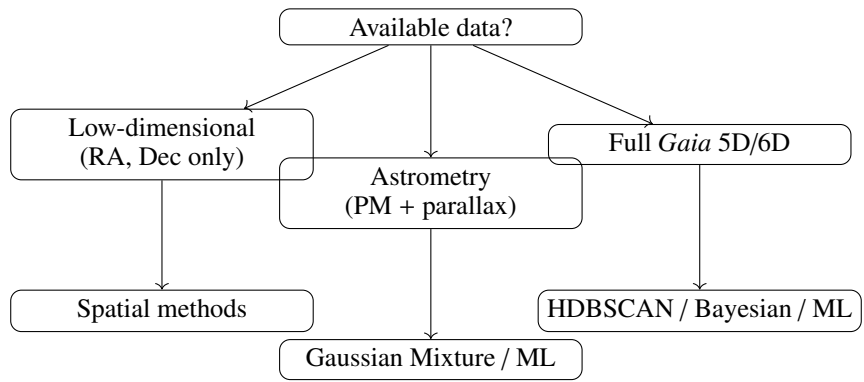
\begin{figure*}
\centering
\begin{tikzpicture}[node distance=2.2cm, every node/.style={draw, rounded corners, align=center, minimum width=4cm}]

\node (a) {Available data?};
\node (b) [below left of=a, xshift=-2cm] {Low-dimensional\\(RA, Dec only)};
\node (c) [below of=a] {Astrometry\\(PM + parallax)};
\node (d) [below right of=a, xshift=2cm] {Full $Gaia$ 5D/6D};

\node (e) [below of=b] {Spatial methods};
\node (f) [below of=c] {Gaussian Mixture / ML};
\node (g) [below of=d] {HDBSCAN / Bayesian / ML};

\draw[->] (a) -- (b);
\draw[->] (a) -- (c);
\draw[->] (a) -- (d);

\draw[->] (b) -- (e);
\draw[->] (c) -- (f);
\draw[->] (d) -- (g);

\end{tikzpicture}

\caption{Decision flowchart for selecting cluster membership determination methods based on available observational data.}
\label{fig:decision_tree}
\end{figure*}

\end{appendix}

\end{document}